\documentclass[twocolumn,showpacs,showkeys,preprintnumbers,amsmath,amssymb]{revtex4}

\usepackage{graphicx}
\usepackage{dcolumn}
\usepackage{bm}

\begin{document}

\title{Coherent bremsstrahlung in a bent crystal}

\author{M.V. Bondarenco}
\email{bon@kipt.kharkov.ua}
\affiliation{%
Kharkov Institute of Physics and Technology, 1 Academic St., 61108
Kharkov, Ukraine.
}%

\date{\today}
\begin{abstract}
Coherent radiation spectrum from high energy $e^\pm$ in a bent
crystal with arbitrary curvature distribution along the longitudinal
coordinate is evaluated, based on the stationary phase
approximation. For a uniformly bent crystal a closed-form expression
for the spectrum is derived. The spectrum features include a dip at
its beginning and the sharp end, which may split into two breaks
depending on the particle incidence angle. Estimates of non-dipole
radiation and multiple scattering effects are given. The value for
the crystal bending angle at which the dipole coherent
bremsstrahlung theory holds best appears to be $\sim
10^{-4}\mathrm{rad}$.
\end{abstract}





\maketitle
\section{Introduction}\label{sec:intro}
Gamma-radiation emitted by electrons and positrons at their
non-channeled passage through planarly oriented bent crystals has
been investigated in a few recent experiments \cite{Afonin,
Scandale} searching for signatures of the charged particle volume
reflection effect \cite{Tar-Vor} in the radiation spectrum. However,
the spectra observed were largely monotonous at typical photon
frequencies $\omega$, and the difference between the measured
spectra from positrons and electrons was basically of order of the
experimental errors. At first sight, that may appear surprising,
since the inherent dynamics and the final reflection angle of volume
reflection itself are known to be sufficiently different in cases of
positively and negatively charged particles.

In article \cite{Chesnokov} dedicated to computer simulation of the
conditions of experiment \cite{Afonin} it was mentioned that the
radiation spectrum must contain a component of so-called coherent
bremsstrahlung in a bent crystal (CBBC) (see \cite{Arutyunov},
appendix), arising at fast charged particle highly over-barrier
motion, when perturbative treatment of particle interaction with the
crystal is valid. Actually, radiation of that type may prove even
dominant when the active crystallographic plane bending angle is
many times larger than the critical value $\theta_c$ \footnote{In
experiments \cite{Afonin, Scandale} it takes on values 3.5
$\theta_c$ and 6 $\theta_c$, which is rather large.} -- then the
particle must spend most of its time traveling at angles to atomic
planes much larger than critical, i. e., flying high above the
potential barrier. Thereat, the frequency of the radiation emitted
by the particle at a given instant is proportional to the local
frequency of atomic plane crossing by the particle, as in ordinary
coherent bremsstrahlung \cite{coh-bremsstr}. In course of the
particle passage, the angle of atomic plane crossing varies, and the
coherent radiation intensity accordingly sweeps over the spectrum.
Close to the volume reflection point, of course, the particle motion
will become non-perturbative, but the radiation from that region
contributes relatively little to the spectrum as a whole, in
contrast to the situation with the elastic scattering, where it is
only the volume reflection point vicinity that contributes to the
particle final deflection angle. The origin of the difference
between the cases is obvious: the magnitude of the intra-crystalline
transverse force is about the same all over the crystal, so all
traversed crystal regions contribute commensurably to the total
irradiation energy. As for the particle deflection angle, it is
sensitive to the force \emph{sign}, and thus receives little
contribution from the regions where the force oscillates rapidly, as
it does at large angles of atomic plane crossing.

In view of the described situation, prior to studies of radiation
features stemming from non-perturbative segments of particle motion
in bent crystals, it seems expedient to determine the shape of
perturbative CBBC spectrum, which yields a wide continuous
background and at the same time provides the conceptually simplest
approximation. Unfortunately, so far it has not been evaluated in a
form suitable for comparison with experiment. It is the purpose of
this article to present a full, though basic calculation, and also
to determine characteristic scales for the physics of the process.
Last not least, we will assess robustness of the simplest CBBC
theory against various deteriorating effects present in nature, such
as multiple scattering and the dipole regime failure. It turns out
that the range of the dipole CBBC theory is rather limited, although
non-vanishing.

In view of the universality of the CBBC radiation in bent crystals
of various shapes (existing examples include cylindrically bent
crystals, sine-shaped bent crystals \cite{Solovyov}, other
microfabricated configurations may appear in future), we extend our
treatment to the case of arbitrary crystal bend profile. Our ability
to cope with it grounds on the applicability of stationary phase
approximation allowing one to treat the crystal curvature as locally
constant. That variant of the stationary phase approximation is of
different origin than the one arising in problems of
synchrotron-like radiation, and does not contradict to the use of
the dipole approximation in radiation.

The paper is organized as follows. In Sec.~\ref{sec:pot-Cart} we
define the bent crystal planar continuous potential and the
corresponding transverse force. In Sec.~\ref{sec:defl} we proceed to
evaluating the particle deflection angle by such a force to the
leading order of high-energy perturbation theory of classical
mechanics. In Sec.~\ref{sec:rad} we evaluate the radiation spectrum
in the dipole approximation, including the quantum effect of
radiation recoil (allowing for photon energies to be of order of the
initial electron's). In Sec.~\ref{sec:conditions} the conditions of
applicability of such an approximation are analyzed.
Sec.~\ref{sec:summary} provides a summary.

\section{Planar continuous potential in a weakly bent crystal}\label{sec:pot-Cart}
\subsection{Geometry definition}
At practice, for the coherent bremsstrahlung at over-barrier
particle passage not to be spoiled by the particle multiple
scattering on the target nuclei it is desirable to work with a
crystal not thicker than a few millimeters (the same situation as
for coherent bremsstrahlung in straight crystals). There are various
techniques for manufacturing such crystals with bent atomic planes
along the short direction, but basically they fall into two
categories. First -- one of the crystal transverse dimensions is
made short, and even shorter than longitudinal (i. e., $\ll1$ mm),
which permits to bend the crystal along the longitudinal direction
\cite{cryst-und}. Second -- both transverse dimensions are made
sufficiently sizeable and the crystal so obtained is bent by some of
the large dimensions, but securing that (short) crystallographic
planes in the crystal, along which the beam is to be sent, acquire
some bending, too. Specifically, the bending of the latter may be
achieved through the action of anti-clastic forces, when the crystal
is deformed simultaneously by \emph{both} large transverse
dimensions with different strength \cite{Ivanov,Guidi} \footnote{In
\cite{Ivanov} it has been asserted that this technique fails with
(110) planar orientation, but works best with (111). Thus, we find
it necessary in our article to discuss cases of (110) and (111) in
parallel.}, or one might just arrange the active planes to be under
some angle $\alpha$ to the large face, and then they must bend along
with the large face, although $\sin\alpha$ times weaker. At the same
time, for issues of particle passage through the crystal the
deviations of the large crystal faces from planes may be neglected.

\begin{figure}
\includegraphics{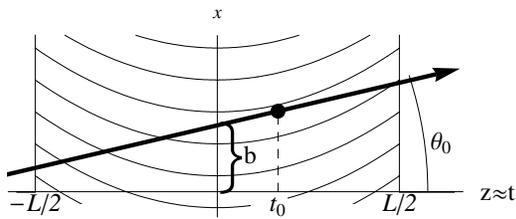}
\caption{\label{fig:angle} Schematic of an ultra-high-energy
particle passage through a thin bent crystal. The vicinity of point
$t_0$ of the trajectory tangency to crystal bent planes gives the
main contribution to the particle deflection angle.}
\end{figure}

In any case, for what concerns description of the particle passage,
the geometry implies particle incidence at some (small) angle
$\theta_0$ to $z$-axis, chosen normal to the crystal large faces
(let the latter be located at positions $z\approx-L/2$ and $z\approx
L/2$)
, and the particle essentially interacts with the continuous
potential of the planes depending only on single coordinate $x$ (see
Fig.~\ref{fig:angle}). The distance between the bent planes is
practically unaffected by the crystal curvature, and the equation
defining each plane takes the form
\begin{equation}\label{}
    x_{\mathrm{pl}}(z)=C_{\mathrm{pl}}+\xi(z),
\end{equation}
constants $C_{\mathrm{pl}}$ being equal-spaced with the inter-planar
distance $d$.

Then, if the continuous inter-planar potential in the bent crystal
was $V_{\mathrm{straight}}(x)$ (a periodic function with period
$d$), which corresponds to an acting force
$F_\mathrm{straight}(x)=-\frac{\partial
V_\mathrm{straight}}{\partial x}$, after bending of this crystal the
force will modify to
\begin{equation}\label{}
    F(x,z)=F_\mathrm{straight} (x-\xi(z))
\end{equation}
(still, it can be regarded as directed along $x$). For  crystals of
constant curvature \footnote{In cases when the active crystal plane
bending is achieved due to secondary elastic (quasimosaic) effects,
the quadratic approximation for $\xi$ may be even more accurate than
that of a circular arc. The author acknowledges communication with
V.~ Guidi on this point.},
\begin{equation}\label{uniform}
    \xi(z)\approx\frac{z^2}{2R},
\end{equation}
with $R=\mathrm{const}$ being the atomic plane bending radius. In
what follows, we will rely on the stationary phase approximation, in
which the crystal curvature is treated locally, and is described by
the local bending radius
\[
    R(z)=\frac1{\left|\xi''(z)\right|},
\]
to emerge naturally in the following.

\subsection{Nearly parabolic continuous potential and the corresponding force}
The dynamics of a high-energy particle in a crystal may be described
by ultra-relativistic classical mechanics \cite{Lindhard}. As we
have agreed, we will use perturbative description of particle
interaction with the crystal; this is a rather common approach in
the theory of coherent bremsstrahlung. Conditions thereof will be
specified later (Sec.~\ref{sec:conditions}).

In the perturbative treatment of classical particle passage
dynamics, as well as in quantum theory, it is advantageous to
express the periodic continuous potential in a form of Fourier
series. Such a representation is economic (provided only a few
lowest harmonics dominate), and at the same time convenient when
proceeding from description of a straight crystal to a bent one. For
evaluation of the particle trajectory and the emitted radiation, of
direct relevance is not the potential but the force acting on the
particle. To define the force -- firstly, in a straight crystal --
it is convenient to choose the origin of $x$-axis in the middle of
some inter-plane interval, with respect to which the potential is an
even function of $x$, whereby the force has to be odd. Then, Fourier
decomposition of the force involves sine functions only:
\[
    F_\mathrm{straight}(x)=\frac2\pi\sum^\infty_{n=1}F_n(-1)^n\sin\frac{2\pi nx}d
\]
(the numerical factors have been introduced for further
convenience).

In the simplest but important case of (110) planar orientation of a
crystal with diamond-type lattice (e. g., silicon), the inter-planar
continuous potential is approximable by a parabola, and the
corresponding force -- by a linear-sawtooth function, whose Fourier
decomposition reads
\begin{subequations}
\begin{eqnarray}
    F^{(110)}_\mathrm{cool}(x)&=&-\frac{2F_{1}}{d}x\big|_{|x|<d/2}+\mathrm{period.}\label{Fsawtooth-spat}\\
    &=&\frac{2F_{1}}\pi\sum^\infty_{n=1}\frac{(-1)^n}n\sin\frac{2\pi
    nx}d.\label{Fsawtooth}
\end{eqnarray}
\end{subequations}
According to (\ref{Fsawtooth-spat}), $F_{1}$ has the meaning of the
force extremal value achieved at $x\to-\frac d2+0$, the sign of
$F_{1}$ being equal to that of the particle charge. On the contrary,
in another important case of orientation (111), there are two
different (but also nearly parabolic) wells within the period of the
continuous potential \cite{Biryukov-UFN}, and the force Fourier
decomposition turns somewhat more complicated:
\begin{eqnarray}\label{F111}
  F^{(111)}_\mathrm{cool}(x)=\frac{32}{\pi d}\sum_{n=1}^\infty \frac{(-1)^n}n\sin\frac{2\pi n
  x}{d}\quad\qquad\qquad\qquad\nonumber\\
  \cdot\left\{\! \left(\frac{V_L}3\!+\!V_S\right)\cos\!\frac{\pi n}4-\frac4{\pi n}\left(\!(-1)^n\frac{V_L}9\!+\!V_S\right)\!\sin\!\frac{\pi n}4
  \right\}
\end{eqnarray}
($V_L$ and $V_S$ have meaning of depths of the alternate unequal
potential wells, while the well widths are exactly $d_L=\frac34d$
and $d_S=\frac14d$). Anyway, once one factors out here the value of
the first Fourier coefficient,
\begin{equation}\label{}
    \left(\frac{V_L}3+V_S\right)\cos\frac{\pi}4+\frac4{\pi
    }\left(\frac{V_L}9-V_S\right)\sin\frac{\pi}4:=\frac{d}{16}F_{1},
\end{equation}
Eq.~(\ref{F111}) will assume the form similar to (\ref{Fsawtooth}):
\begin{eqnarray}\label{Fsawtooth111}
    F^{(111)}_\mathrm{cool}(x)=\frac{2F_{1}}\pi\sum^\infty_{n=1}\frac{(-1)^nc_n}n\sin\frac{2\pi
    nx}d.
\end{eqnarray}
Here $c_n$ is a sequence of coefficients of order unity, neither
increasing nor decreasing as $n\to\infty$, and by definition
$c_1=1$. \footnote{Taking $V_L \approx 26.5\,\mathrm{eV}$, $V_S
\approx 7.5\,\mathrm{eV}$, one obtains $c_2=-0.9$, $c_3=-1.7$,
$c_4=-2.2$, $c_5=-1.4$.}

\begin{figure}
\includegraphics{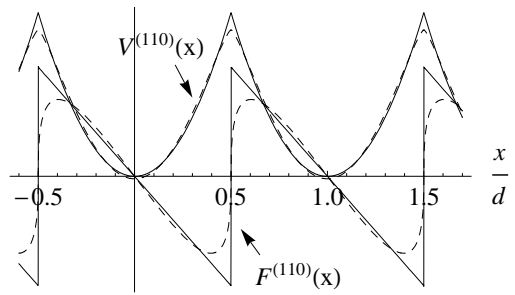}
\caption{\label{fig:sawtooth} Inter-planar continuous potential and
the corresponding force (Eq.~(\ref{Ftherm})) shapes (in arbitrary
units) for positively charged particles in a silicon crystals with
(110) orientation. Solid lines: $\epsilon=0$ (cooled crystal,
Eq.~(\ref{Fsawtooth-spat})); dashed lines: $\epsilon=0.4$
(room-temperature crystal, Eq.~(\ref{Ftherm})). For negatively
charged particles the signs of the functions reverse.}
\end{figure}

To take into account thermal smearing of the potential, i. e., the
force continuity at the locations of atomic planes, the simplest
though heuristic trick is to increase the power of $n$ in the
overall $\frac 1n$ factor of the trigonometric series:
\begin{eqnarray}\label{Ftherm}
    F_\mathrm{therm}(x)=\frac{2F_{1}}\pi\sum^\infty_{n=1}\frac{(-1)^nc_n}{n^{1+\epsilon}}\sin\frac{2\pi
    nx}d,\\
    \epsilon=\epsilon(T)>0.\quad\qquad\qquad\qquad\nonumber
\end{eqnarray}
At that, the sequence $c_n$ (or its parameters $V_L$, $V_S$) may
need to be corrected, but still the series is dominated by the first
term, for which $c_1\approx1$. Such a modification acts similarly to
the conventional Debye-Waller exponential factor (which, in
principle, is also heuristic, only its first order Maclaurin term
being rigorously related to thermal averages). We refrain here from
discussing the exact relation of $\epsilon$ with temperature $T$,
only indicate that for the case of Si (110) at room temperature
agreement with the potentials used in the literature is achieved at
$\epsilon\approx0.4$ (see Fig.~\ref{fig:sawtooth}), whereas for Si
(111) it takes $\epsilon\sim1$. To supply more motivation to our
ansatz, note that in what follows the summation of series of the
type (\ref{Ftherm}) with constant $\epsilon$ and simple $c_n$ will
yield Riemann $\zeta$-related functions. Such functions emerge as
well for a zero-temperature potential ($\epsilon=0$), only in the
latter case the function arguments being integer or half-integer.
Our approach corresponds to extension of those arguments to
arbitrary fractional values, i. e. to an ``analytic continuation",
in order to model the effect of the temperature in a simplest way.
None of the following numerical results (serving as estimates)
depends crucially on this technique.

Practical bent crystals are usually manufactured from silicon. The
relevant physical parameters for silicon are
\begin{subequations}\label{Fd}
\begin{eqnarray}
    d=1.9\AA, \quad
    |F_{1}|\approx6\frac{\mathrm{GeV}}{\mathrm{cm}},\qquad (\mathrm{Si}\,(110))\label{Tar-pars}\\
    d=3.1\AA, \quad
    |F_{1}|\approx4\frac{\mathrm{GeV}}{\mathrm{cm}}.\qquad (\mathrm{Si}\,(111))\label{111-pars}
\end{eqnarray}
\end{subequations}
Note that product $|F_{1}|d$ for those cases has practically
identical values, which is important for the subsequent numerical
estimates. But all our figures will be drawn only for a simpler case
(110).

\section{Infinitesimal deflection angle}\label{sec:defl}

In this section we shall analyze the elastic scattering of
ultra-high energy particles by the continuous force defined in the
previous section. Choose the time reference point at the moment of
the particle passage through the middle of the crystal ($z=0$), so
that we may equate $t\simeq z$ (we will use units $c=\hbar=1$).
Since the beam width in practice is always greater than the
inter-planar distance, there is essentially a uniform distribution
of particles in transverse impact parameters. Defining impact
parameter $b$ of an individual particle as the trajectory initial
asymptote intercept on $x$-axis, i. e. at $z=0$ (see
Fig.~\ref{fig:angle}), the force acting on the particle in a bent
crystal can be written as
\begin{eqnarray}\label{force-f-L}
    F(t)&=&\Theta\left(\frac
    L2-|t|\right)\frac{2F_{1}}\pi\nonumber\\
    &\cdot&\sum_{n=1}^\infty
    \frac{(-1)^n}{n^{1+\epsilon}}\sin\left(2\pi
    n\frac{b+\theta_0t-\xi(t)}d\right),\qquad
\end{eqnarray}
with $\Theta(v)$ -- the Heavyside unit step function (zero for
negative arguments and unity for positive ones).

First of all, let us evaluate the particle deflection angle.
Asymptotically, to leading order in the potential to energy ratio
$V/E$, the deflection angle is proportional to the integral of force
(\ref{force-f-L}) along the particle unperturbed straight path
\footnote{Thereby, it can also be regarded as first Born
approximation in coupling with the crystalline field.}:
\begin{eqnarray}\label{th1}
    \theta_{\mathrm{Born}}\left(\theta_0,b\right)&=&\frac1E\int_{-L/2}^{L/2}F(t)dt\nonumber\\
    &=&\frac{2}{\pi R_c}\sum_{n=1}^\infty
    \frac{(-1)^nc_n}{n^{1+\epsilon}}\nonumber\\
    &\cdot&\int_{-L/2}^{L/2}\sin\left(2\pi
    n\frac{b+\theta_0t-\xi(t)}d\right)dt,\qquad
\end{eqnarray}
where
\begin{equation}\label{Rc}
    R_c=E/F_{1}
\end{equation}
is the Tsyganov critical radius \cite{Tsyganov} (the above
definition of $R_c$ yet allows it, along with $F_{1}$, to have
different sign depending on the particle charge sign). If the
crystal bending is macroscopic, in the sense that displacement $\xi$
of the planes is (generally) $\gg d$, the integrand is rapidly
oscillatory. For evaluation of such an integral, one may employ the
stationary phase approximation \cite{Olver}. This requires, in the
first place, finding stationary phase points $t_s$ at which
\begin{equation}\label{}
    t_s(\theta_0):\qquad \theta_0-\frac{d\xi}{dt}\bigg|_{t=t_s}=0,
\end{equation}
i. e., the points of tangency of a ray with slope $\theta_0$ to the
family of bent crystalline planes. For a convex function $\xi$ such
a point is unique -- and for simplicity we will assume this to be
the case, dubbing it $t_0$ (see Fig.~\ref{fig:angle}). Then,
expanding function $\xi(t)$ in Taylor series about $t_0$ up to
quadratic terms, one brings (\ref{th1}) to the form
\begin{eqnarray}\label{stat-phase}
    \theta_{\mathrm{Born}}\left(\theta_0,b\right)\approx\frac{2}{\pi R_c}\sum_{n=1}^\infty
    \frac{(-1)^nc_n}{n^{1+\epsilon}}\quad\qquad\qquad\nonumber\\
    \cdot\Im\mathfrak{m}\int_{-L/2}^{L/2}dt\exp\bigg\{i\frac{2\pi
    n}d\Big(b+\theta_0t_0-\xi(t_0)\quad\nonumber\\
    -\frac12\xi''(t_0)(t-t_0)^2\Big)\bigg\}.\quad
\end{eqnarray}

\begin{figure}
\includegraphics{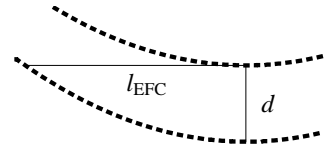}
\caption{\label{fig:l-EFC} Geometric interpretation of the coherence
length of a fast particle traversing a bent crystal:
$l_{\mathrm{EFC}}$ is a half-chord within a curved crystalline plane
(radius $R$) tangential to the next curved plane at distance $d$
from the initial one.}
\end{figure}

Now, if point $t_0$ belongs to the interval $-\frac L2<t_0<\frac
L2$, the integral converges in a small vicinity of this point of the
width
\begin{equation}\label{coh-length}
    \left|t-t_0\right|\sim
    l_{\mathrm{EFC}}(t_0)=\sqrt{2R\left(t_0\left(\theta_0\right)\right)d},
\end{equation}
where
\begin{equation}\label{Rloc}
    R(t)=\frac1{\left|\xi''(t)\right|}
\end{equation}
(for a geometric construction see Fig.~{\ref{fig:l-EFC}}).
Physically, this is the length on which the deflecting external
field acts periodically, and will be referred to as the external
field coherence (EFC) length. Then, the integration limits in
(\ref{stat-phase}) may as well be extended to infinity, and the
integral evaluates by a standard formula
\[
    \int^\infty_{-\infty}e^{iAt^2}dt=e^{i\frac\pi 4
    \mathrm{sgn}A}\sqrt{\frac\pi{|A|}},\qquad (\Im\mathfrak{m} A=0)
\]
with $\mathrm{sgn}$ function defined as $\mathrm{sgn}A=-1(+1)$ if
$A<0(A>0)$. If, on the contrary, $t_0$ falls beyond the integration
interval, the integrand is everywhere rapidly oscillatory, and the
result is small (yet there are contributions from the end-points,
inferior to those from stationary phase points, but they will be
neglected throughout for simplicity). With this accuracy,
\begin{eqnarray}\label{theta}
    \theta_{\mathrm{Born}}\left(\theta_0,b\right)\approx\Theta\left(\frac L2-|t_0(\theta_0)|\right)\frac{\!\sqrt{R\left(t_0\left(\theta_0\right)\right)d}}{R_c}\quad\nonumber\\
    \cdot\frac2\pi\sum_{n=1}^\infty
    \frac{(-1)^n c_n}{n^{3/2+\epsilon}}\sin\Bigg(-\frac\pi4\mathrm{sgn}\xi''\left(t_0(\theta_0)\right)\qquad\qquad\nonumber\\
    \quad+2\pi
    n\left[\frac{b+\theta_0t_0(\theta_0)-\xi\left(t_0(\theta_0)\right)}{d}\right]\Bigg).
\end{eqnarray}
Hence, the magnitude of deflection angles is determined by the
crystalline plane curvature radius in the point of the trajectory
tangency to the bent planes.

As for functions $t_0(\theta_0)$, $\xi\left(t_0(\theta_0)\right)$
appearing in (\ref{theta}), they may be obtained explicitly only if
$\xi(t)$ is a sufficiently simple analytic function. For instance,
in case of a crystal of constant curvature (\ref{uniform}) they
express through the only available parameter -- the plane bending
radius:
\begin{eqnarray}
    t_0=R\theta_0,\quad \xi(t_0)=\frac R2\theta_0^2,\quad \theta_0t_0-\xi(t_0)=\frac
    R2\theta_0^2.\nonumber\\
    \qquad (R=\mathrm{const})\qquad\qquad\qquad\qquad\nonumber
\end{eqnarray}
So, at practice reversible analytic parameterizations of the bending
profile $\xi(t)$ are favored.

For the case of orientation (110), when $c_n\equiv1$, the sum in the
right-hand side of (\ref{theta}) can be expressed in terms of
Hurwitz (generalized Riemann) $\zeta$-functions:
\begin{eqnarray}\label{zeta}
    \frac2\pi\sum_{n=1}^\infty
    \frac{(-1)^n}{n^{3/2+\epsilon}}\sin\left\{-\frac\pi4\mathrm{sgn}\xi''(t_0)+2\pi
    n{\beta}\right\}\qquad\quad\nonumber\\
    =\mathrm{sgn}\xi''(t_0)\frac{2(2\pi)^{\frac12+\epsilon}}{\Gamma\left(\frac32+\epsilon\right)}\qquad\qquad\qquad\quad\nonumber\\
    \cdot\Bigg(\cos\!\frac{\pi\epsilon}2\zeta\!\left(-\frac12-\epsilon,\,\left\{\frac12+{\beta}\mathrm{sgn}\xi''(t_0)\right\}_{\mathrm{f}}\right)\,\nonumber\\
    -\sin\!\frac{\pi\epsilon}2\zeta\!\left(-\frac12-\epsilon,\,\left\{\frac12-{\beta}\mathrm{sgn}\xi''(t_0)\right\}_{\mathrm{f}}\right)\!\Bigg),
\end{eqnarray}
where
\begin{equation}\label{beta}
    {\beta}\left(\frac
bd,\theta_0\right)=\frac{b+\theta_0t_0(\theta_0)-\xi\left(t_0(\theta_0)\right)}d
\end{equation}
characterizes the impact parameter of an oblique trajectory relative
to the bent planes in point $z=t_0$, and $\zeta(\alpha,v)$ is the
Hurwitz zeta-function \cite{Apostol}, while braces
$\{\ldots\}_{\mathrm{f}}$ in its second argument indicate the entry
fractional part (ranging from 0 to 1). For orientation (111) the
result can be expressed through Hurwitz zeta-functions in a similar
manner, but it is somewhat more bulky and shall not be quoted
herein.

\begin{figure}
\includegraphics{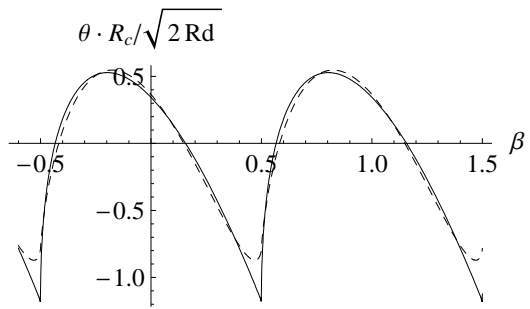}
\caption{\label{fig:zeta} Deflection angle of a positively charged
particle in a bent Si (110) crystal, in units of
$\frac{\sqrt{2Rd}}{R_c}$, vs. the impact parameter variable
$\beta=\frac bd+\mathrm{const}$ (see Eq.~(\ref{beta})). Solid line:
$\epsilon=0$ (cooled crystal), dashed line: $\epsilon=0.4$
(room-temperature crystal).}
\end{figure}

Function (\ref{zeta}) (shown in Fig.~\ref{fig:zeta}) is not
particularly sensitive to the value of $\epsilon$, except around the
fracture points. The latter ones are located at
\begin{equation}\label{frac-points}
    \beta=\pm\frac12,\pm\frac32,\, \ldots\, .
\end{equation}
In those points, function (\ref{zeta}) (and therewith (\ref{theta}))
is extremal and achieves the value
\begin{eqnarray*}
  \max_b\frac{\theta_{\mathrm{Born}}(\theta_0,b)}{\mathrm{sgn}\xi''}=-\Theta\left(\frac
    L2-|t_0(\theta_0)|\right)\qquad\qquad\nonumber\\
  \cdot\zeta\left(\frac32+\epsilon\right)\!\frac{\!\sqrt{2R\left(t_0\left(\theta_0\right)\right)d}}{\pi
    R_c},
\end{eqnarray*}
where $\zeta(\alpha)=\sum_{n=1}^{\infty}n^{-\alpha}$ is the ordinary
Riemann zeta-function.

The noticeable asymmetry of function(s) in Fig.~\ref{fig:zeta} can
be traced to the phase shift $\frac\pi4$, arising within the
stationary phase approximation. The average value of the deflection
angle over the impact parameters is strictly zero, as an average of
a sum of sine functions over their full period. The physical reason
behind that is the uniform distribution of an unperturbed particle
flow over the crystal, entailing equal influence of positive and
negative forces on the entire beam. So, in the adopted first order
of perturbation theory there is no signature of volume reflection of
the beam. Non-zero, however, is the angular spread acquired by the
beam, whose measure is the deflection angle mean square:
\begin{eqnarray}\label{meanthetasquare}
    \left\langle\theta_{\mathrm{Born}}^2\right\rangle&=&\frac1d\int_0^ddb\theta^2(\theta_0,b)\nonumber\\
    &\approx&\frac{2R\left(t_0(\theta_0)\right)d}{\pi^2R_c^2}\sum_{n=1}^\infty\frac{c_n^2}{n^{3+2\epsilon}}.
\end{eqnarray}
The latter quantity will also play part in the treatment of
radiation.

\section{Dipole coherent bremsstrahlung}\label{sec:rad}
Having established in the previous section the description of the
particle passage through the crystal, we are in a position to
calculate the accompanying radiation. As is typical for
bremsstrahlung from relativistic particles, the radiation is
concentrated within a cone of angles $\sim\gamma^{-1}$ about the
forward direction, and those small angles may be treated
inclusively. Let $\omega$ stand for the photon frequency (energy).
Within the dipole approximation, the spectrum of radiation
integrated over emission angles, and averaged over the impact
parameters $b$ of particles in the beam, expresses through the
acting force as \cite{BKStr,coh-bremsstr}
\begin{eqnarray}\label{dEdomega-dip}
    \frac{d\!E_{\mathrm{CBBC}}}{d\omega}\!=\!\frac{e^2\!E'\omega}{2\pi
    E^3}\!\int^\infty_{q_{\mathrm{min}}}\!\frac{dq}{q^2}\left(1\!+\!\frac{\omega^2}{2EE'}\!-\!\frac{2q_{\mathrm{min}}}{q}\!+\!\frac{2q^2_{\mathrm{min}}}{q^2}\right)\nonumber\\
    \cdot\frac1d\int_0^ddb\left|F\left(q,\theta_0,b\right)\right|^2,\quad
\end{eqnarray}
where
\begin{equation}\label{qmin-def}
     E'=E-\omega,\qquad q_{\mathrm{min}}=q_{\mathrm{min}}(\omega)=\frac{\omega m^2}{2EE'},
\end{equation}
(allowing for $\omega\sim E$, to account for quantum radiation
recoil effects), and
\begin{equation}\label{Fqdef}
    F\left(q,\theta_0,b\right)=\int^\infty_{-\infty}dte^{iqt}F\left(t,\theta_0,b\right).
\end{equation}
Thus, $q$ is a frequency of the force acting on the particle, and
(\ref{Fqdef}) is the Fourier transformation of the intra-crystal
force.

With $F(t)$ given by Eq.~(\ref{force-f-L}), its Fourier transform is
conveniently evaluated by decomposing the sine function into a pair
of exponentials:
\begin{eqnarray}\label{Fq}
    F(q,\theta_0,b)=\frac{2F_{1}}\pi\sum^\infty_{n=1}\frac{(-1)^nc_n}{n^{1+2\epsilon}}\qquad\qquad\qquad\qquad\nonumber\\
    \cdot\int^{L/2}_{-L/2}dte^{iqt}\sin\left(2\pi
    n\frac{b+\theta_0t-\xi(t)}d \right)\quad\nonumber\\
    =\!\frac{F_{1}}{\pi i}\!\sum^\infty_{n=1}\frac{(-1)^n}{n^{1+\epsilon}}\!\!\int^{L/2}_{-L/2}dt\Bigg(e^{i\left(2\pi n\frac{b+\theta_0t-\xi(t)}d+qt\right)}\quad\nonumber\\
    -e^{-i\left(2\pi
    n\frac{b+\theta_0t-\xi(t)}d-qt\right)}\Bigg).\quad
\end{eqnarray}
When averaging the square of (\ref{Fq}) over the impact parameters,
we employ the identity
\begin{equation}\label{ident-delta}
    \frac 1d\int^d_0dbe^{2\pi in\frac bd}e^{-2\pi im\frac
    bd}=\delta_{nm},
\end{equation}
by virtue of which
\begin{eqnarray}\label{averF}
    \frac1d\int_0^ddb\left|F(q,\theta_0,b)\right|^2\qquad\qquad\qquad\qquad\qquad\qquad\,\nonumber\\
    =\frac{F^2_{1}}{\pi^2}\sum^\infty_{n=1}\frac{c_n^2}{n^{2+2\epsilon}}\Bigg(\!\left|\int^{L/2}_{-L/2}dte^{i\left(2\pi n\frac{\theta_0t-\xi(t)}d+qt\right)}\right|^2\quad\nonumber\\
    +\left|\int^{L/2}_{-L/2}dte^{i\left(2\pi
    n\frac{\theta_0t-\xi(t)}d-qt\right)}\right|^2\Bigg)\quad
\end{eqnarray}
(note the absence of interference between the exponents after the
averaging).

Evaluation of each of the two oscillatory integrals in (\ref{averF})
is carried out by the stationary phase approximation, as in the
previous section. The external field coherence length is the same as
(\ref{coh-length}), only the location of the stationary-phase point
$t_{n\pm}$, about which function $\xi(t)$ has to be Taylor-expanded,
now depends on $\theta_0$ and $q/n$. The equations for stationary
phase points read
\begin{equation}\label{t}
    t_{n\pm}(q,\theta_0):\quad\xi'\left(t_{n\pm}\right)-\theta_0=\pm\frac{qd}{2\pi
    n}.
\end{equation}
(In accord with (\ref{averF}), radiation from different stationary
phase points does not interfere). Physically, $\xi'-\theta_0$
represents the angle between the beam and the crystalline planes in
the stationary phase point. Thereby, Eq.~(\ref{t}) may be viewed as
a local coherent bremsstrahlung condition, in which the local
frequency of the driving external force is proportional to the local
frequency of crystalline plane crossing by the particle, which in
turn is proportional to the local angle of the trajectory
inclination to the planes (cf.~\cite{coh-bremsstr}). Ultimately, the
approximate $t$-integration gives
\begin{eqnarray}\label{pm}
    \left|\int^{L/2}_{-L/2}dte^{i\left(2\pi
    n\frac{\theta_0t-\xi(t)}d\pm
    qt\right)}\right|^2\qquad\qquad\qquad\qquad\nonumber\\
    \approx\Theta\left(\frac
    L2-\left|t_{n\pm}(q,\theta_0)\right|\right)\frac{R\left(t_{n\pm}(q,\theta_0)\right)d}n.
\end{eqnarray}

\begin{figure}
\includegraphics{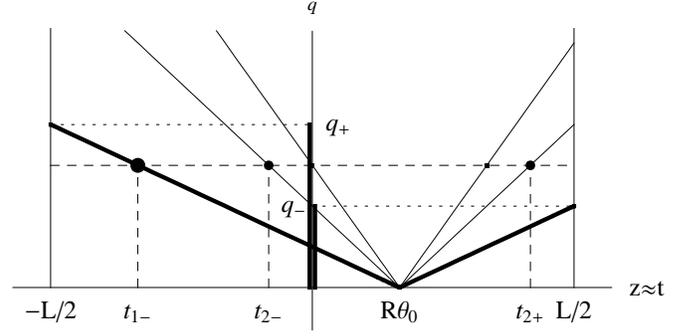}
\caption{\label{fig:freq} Local frequencies of plane crossing (thick
oblique lines), and the frequencies of higher harmonics (thinner
oblique lines), as functions of the particle longitudinal coordinate
(time). Drawn for a specific case of crystal with constant
curvature, when $q(t)$ dependencies are linear. Vice versa, the
construction may be used for determination of stationary phase
points for a given frequency (dashed lines). Thick intercepts on the
$q$-axis schematically show a double-step distribution of the force
frequencies described by function $|F(q)|^2$ (see also
Fig.~\ref{fig:Fspectrum}c).}
\end{figure}

Substitution of (\ref{averF}, \ref{pm}) to (\ref{dEdomega-dip})
leads to the result for the radiation spectrum of coherent
bremsstrahlung in a bent crystal:
\begin{eqnarray}\label{dECBBC-gen}
    \frac{dE_{\mathrm{CBBC}}}{d\omega}\approx\frac{e^2F^2_{1}d}{2\pi^3}\frac{E'\omega}{E^3}\quad\qquad\qquad\qquad\qquad\qquad\quad\nonumber\\
    \cdot\sum^\infty_{n=1}\frac{c_n^2}{n^{3+2\epsilon}}\int^\infty_{q_{\mathrm{min}}}\frac{dq}{q^2}\left(1+\frac{\omega^2}{2EE'}-\frac{2q_{\mathrm{min}}}{q}+\frac{2q^2_{\mathrm{min}}}{q^2}\right)\nonumber\\
    \cdot\bigg\{\Theta\left(\frac L2-\left|t_{n+}(q,\theta_0)\right|\right)R\left(t_{n+}(q,\theta_0)\right)\qquad\nonumber\\
    +\Theta\left(\frac
    L2-\left|t_{n-}(q,\theta_0)\right|\right)R\left(t_{n-}(q,\theta_0)\right)\!\bigg\}.\quad
\end{eqnarray}
By virtue of the power factor $\frac1{n^{3+2\epsilon}}$, this sum is
strongly dominated by the term $n=1$, so the coherent bremsstrahlung
spectrum shapes for crystal orientations (110) and (111) appear to
be only marginally different, and the temperature effect on the
coherent radiation is also small.

In what follows we will concentrate on application of formula
(\ref{dECBBC-gen}) to crystals of constant curvature. In that case
the solution to Eq.~(\ref{t}) is
$t_{n\pm}=\left(\theta_0\pm\frac{qd}{2\pi n}\right)\!R$ (see
Fig.~\ref{fig:freq}), and in (\ref{dECBBC-gen}) one may draw
constant $R$ out of the integral, which allows one to accomplish the
integration in terms of elementary functions. Introducing parameters
\begin{equation}\label{q-pm}
    q_\pm=\frac{2\pi}d\left(\frac L{2R}\pm|\theta_0|\right)
\end{equation}
(signifying the frequencies of active crystalline plane crossing at
the entrance and at the exit from the crystal) and a function
\begin{subequations}
\begin{eqnarray}
    D\left(v,\frac\omega E\right)&=&\int_v^1dy\left(1+\frac{\omega^2}{2EE'}-2y+2y^2\right)\quad\qquad\nonumber\\
    &=&(1-v)\left(\frac{2-v+2v^2}3+\frac{\omega^2}{2EE'}\right)\,
    \label{D-def1}\\
    &\equiv&
    \frac13+\frac12\left(\frac12-v\right)+\frac23\left(\frac12-v\right)^3\nonumber\\
    &\,&+\left(1-v\right)\frac{\omega^2}{2EE'}, \label{D-def}
\end{eqnarray}
\end{subequations}
\[
(0\leq
    v\leq1)
\]
the expression for the radiation spectrum assumes the form
\begin{eqnarray}\label{dECBBC-unif}
    \frac{dE_{\mathrm{CBBC}}}{d\omega}=\frac{e^2F^2_{1}Rd}{\pi^3
    m^2}\frac{E'^2}{E^2}\qquad\qquad\qquad\qquad\qquad\qquad\nonumber\\
   \cdot\sum^\infty_{n=1}\frac{c_n^2}{n^{3+2\epsilon}}\bigg\{\Theta\left(nq_--q_{\mathrm{min}}\right)D\left(\frac{q_{\mathrm{min}}}{nq_-},\frac\omega E\right)\quad\qquad\qquad\nonumber\\
    +\Theta(nq_-+q_{\mathrm{min}})\Theta(nq_+-q_{\mathrm{min}})D\left(\frac{q_{\mathrm{min}}}{nq_+},\frac\omega E\right)\quad\qquad\,\nonumber\\
    +\Theta(-nq_-\!-\!q_{\mathrm{min}})\!\left[D\!\left(\frac{q_{\mathrm{min}}}{nq_+},\frac\omega E\right)\!-\!D\!\left(\frac{q_{\mathrm{min}}}{n|q_-|},\frac\omega
    E\right)\right]\!\!\bigg\}.\nonumber\\
\end{eqnarray}

\begin{figure}
\includegraphics{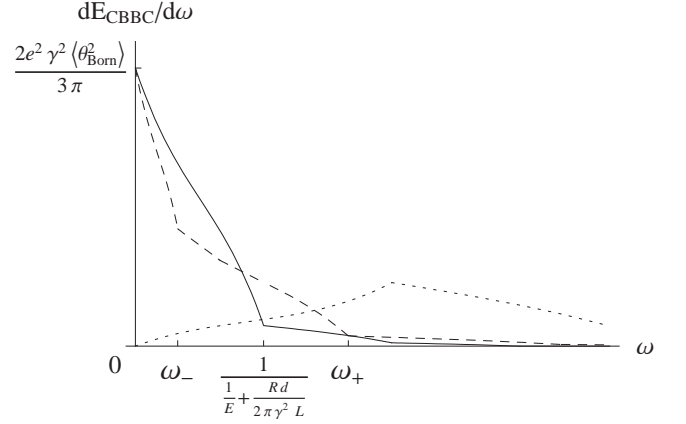}
\caption{\label{fig:spectrum} Coherent radiation spectra (neglecting
multiple scattering and incoherent bremstrahlung), for a fixed value
of $\frac{2\pi L\gamma^2}{Rd}\ll E$ and several values of
$|\theta_0|$. Solid line: $\theta_0=0$ ($\omega_-$ and $\omega_+$
coincide); dashed: $|\theta_0|=\frac L{3R}$; dotted:
$|\theta_0|=\frac LR$. The impact of temperature effects on
radiation is negligible; the type of crystal orientation ((110) or
(111)) mainly affects the frequency and intensity scales.}
\end{figure}
\begin{figure}
\includegraphics[width=85mm]{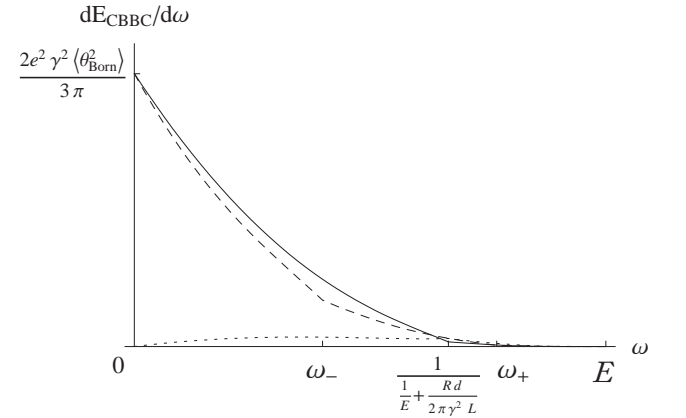}
\caption{\label{fig:spectrum-he} Same as Fig.~\ref{fig:spectrum},
for $\frac{2\pi L\gamma^2}{Rd}=2E$. At this electron energy, the
radiation spectrum shape is already strongly influenced by the
quadratic factor $E'^2/E^2$ in Eq.~(\ref{dECBBC-unif}), obscuring
the spectrum features. }
\end{figure}

The behavior of the spectrum for different incidence angles is
illustrated in Figs.~\ref{fig:spectrum}, \ref{fig:spectrum-he}.
Despite being composed of discontinuous $\Theta$-functions, in total
(\ref{dECBBC-unif}) is everywhere continuous, as is conditioned by
its initial integral representation (\ref{dECBBC-gen}). Still, there
are discontinuities in the derivative of (\ref{dECBBC-unif})
manifesting themselves as sharp curve breaks (``ankle"-type). Beyond
the first two, major breaks \footnote{They coalesce into one in the
limit $\theta_0\to 0$.} (corresponding to $n=1$) the spectrum
effectively ends, and only contributions from higher harmonics
remain. Those main ``ankles" are located at photon energies
\footnote{Eq.~(\ref{omega-pm}) results as a solution to equation
$q_{\min}(\omega_{\pm})\equiv\frac{m^2\omega_{\pm}}{2E(E-\omega_{\pm})}=|q_{\pm}|$.}
\begin{equation}\label{omega-pm}
  \omega_\pm=\frac1{\frac1E+\frac1{2\gamma^2|q_\pm|}}.
\end{equation}
Note that when $|\theta_0|$ is only slightly below $\frac{L}{2R}$,
then $\omega_-\ll\omega_+$, and at $\omega\leq\omega_-$ the spectrum
develops a sharp spike, although superimposed on the background of
equal height. That condition corresponds to a trajectory nearly
tangential to the crystalline planes at the entrance or exit from
the crystal, and although this feature may be of experimental
utility, one should beware that our present stationary phase
approximation neglecting end-point effects, as well as the dipole
approximation itself, are in substantial error there.

There are other important features of the CBBC spectrum concerning
dependencies on the geometric parameters $R$, $L$ and $\theta_0$:

\begin{description}
  \item[(i)] It is natural that the differential cross-section of coherent
radiation is proportional to the square of the field strength and to
the square of the coherence length (\ref{coh-length}). We accentuate
that the coherence length in our problem is of external origin and
\emph{independent of} $\omega$ -- that is a length on which the
particle-crystal interaction may be regarded as periodic. In
contrast, the photon formation length \footnote{Which is sometimes
also referred to as coherence length (in the sense of coherence in
the process of photon emission), hence possible confusion.}
$l_{\mathrm{form}}=q^{-1}_{\min}(\omega)\gtrsim q^{-1}_\pm$ depends
on $\omega$. It sets the scale of resulting photon energies,
correlating with spacings $q^{-1}_\pm$ between the neighboring
crystalline planes measured along the particle trajectory at the
entrance to and at the exit from the crystal.
  \item[(ii)] One can check that as $\omega\to0$ the limit of (\ref{dECBBC-gen}) is
\begin{equation}\label{soft limit}
    \frac{dE_{\mathrm{CBBC}}}{d\omega}\underset{\omega\to0}\to\frac{2e^2}{3\pi}\gamma^2\left\langle\theta^2_{\mathrm{Born}}\right\rangle
\end{equation}
with $\left\langle\theta^2_{\mathrm{Born}}\right\rangle$ given by
Eq.~(\ref{meanthetasquare}). Apparently, this value does not depend
on the crystal thickness $L$.
  \item[(iii)] The total radiation energy emitted per one electron
  \begin{equation}\label{}
    E_{\mathrm{CBBC}}=\int_0^Ed\omega\frac{dE_{\mathrm{CBBC}}}{d\omega}
  \end{equation}
  expresses rather simply and quite differently in two limiting cases: when the photon recoil effects are negligible, and when they
  are crucial. If the ``moderately high energy" condition $2\gamma^2q_+\ll E$
 is met, then $q_{\mathrm{min}}\approx\frac\omega{2\gamma^2}$ and the
second argument of all the $D$-functions in (\ref{dECBBC-unif}) may
be put to zero. For this case, one finds
\begin{eqnarray}
  E_{\mathrm{CBBC}} \simeq L\frac{8e^2}{3\pi^2}\gamma^2\frac{F^2_{1}}{m^2}\sum^\infty_{n=1}\frac{c_n^2}{n^{2+2\epsilon}}.\\
  \left(\gamma\ll\frac {md}{2\pi}\frac RL\sim10^2\frac
  RL\right)\qquad\label{soft-rad-cond}
\end{eqnarray}
In contrast to the differential intensity, the total emitted energy
here is proportional not to the square of the coherence length but
to the crystal thickness. Remarkably, it does not depend on $R$, nor
$\theta_0$, and just equals to the total energy of coherent
bremsstrahlung radiation in a \emph{straight} crystal of thickness
$L$.

If the opposite condition $2\gamma^2q_+\gg E$ is realized, then the
first argument of all the $D$-functions in (\ref{dECBBC-unif}) may
be put to zero, giving
\begin{eqnarray}\label{}
    E_{\mathrm{CBBC}}&\simeq&\frac{dE_{\mathrm{CBBC}}}{d\omega}\bigg|_{\omega\to0}\int_0^{E}d\omega\frac{E'^2}{E^2}\!\left(1\!+\!\frac{3\omega^2}{4EE'}\!\right)\nonumber\\
    &=&E\frac{19e^2}{36\pi^3}\frac{F^2_{1}Rd}{m^2}\!\sum^\infty_{n=1}\!\frac{c_n^2}{n^{3+2\epsilon}}\Theta\!\left(\frac
    L{2R}-|\theta_0|\right)\!.\qquad
\end{eqnarray}
\begin{equation}\label{hard-rad-cond}
    \left(\gamma\gg\frac {md}{2\pi}\frac RL\sim10^2\frac
  RL\right)
\end{equation}

So, in this limit even the total radiation energy does not depend on
the crystal thickness.
    \item[(iv)] At
\begin{equation}\label{straight-limit}
    |\theta_0|\gg \frac L{2R}
\end{equation}
(a large incidence angle or the straight crystal limit),
Eqs.~(\ref{q-pm}, \ref{omega-pm}) yield $q_-\simeq-q_+<0$,
    $\omega_-\simeq\omega_+$. Then in (\ref{dECBBC-unif}) the term
    with $\Theta\left(nq_--q_{\mathrm{min}}\right)$ vanishes. The
    next term containing
    $\Theta(nq_-+q_{\mathrm{min}})\Theta(nq_+-q_{\mathrm{min}})$ is
    non-zero only in relatively small intervals $n|q_-|\leq
    q_{\mathrm{min}}<
    nq_+$, yet the corresponding $D$-function has its first argument close to
    unity and thereby is small (cf. Eq.~(\ref{D-def1})):
\begin{equation}\label{}
    D\left(\frac{q_{\mathrm{min}}}{nq_+},\frac\omega E\right)\simeq
    \left(1-\frac{q_{\mathrm{min}}}{nq_+}\right)\!\left(1+\frac{\omega^2}{2EE'}\right)\ll1.
\end{equation}
Lastly, the term containing $\Theta(-nq_- -q_{\mathrm{min}})$
contributes on the entire interval $q_{\mathrm{min}}\leq n|q_-|$, i.
e., basically at $\omega\leq\omega_-\simeq\frac1{\frac1E+\frac
d{4\pi\gamma^2|\theta_0|}}$, but there is a valuable cancelation
between the corresponding $D$-functions:
\begin{eqnarray}\label{}
    D\left(\frac{q_{\mathrm{min}}}{nq_+},\frac\omega
E\right)-D\left(\frac{q_{\mathrm{min}}}{n|q_-|},\frac\omega
    E\right)\quad\qquad\qquad\qquad\nonumber\\
    \simeq \frac{\partial}{\partial v}D\left(v,\frac\omega
E\right)\Big|_{v=\frac{q_{\mathrm{min}}}{n|q_-|}}\frac{q_{\mathrm{min}}}n\left(\frac1{q_+}-\frac1{|q_-|}\right)\qquad\nonumber\\
\cong\left(1-2\frac{q_{\min}}{n|q_-|}+2\frac{q^2_{\min}}{n^2|q_-|^2}+\frac{\omega^2}{2EE'}\right)\frac{q_{\min}}{nq^2_-}\frac{2\pi
L}{Rd},
\end{eqnarray}
\[
|q_-|\simeq\frac{2\pi|\theta_0|}d.
\]
Therewith, the radiation spectrum reduces to
\begin{eqnarray}\label{rr}
    \frac{dE}{d\omega}\simeq L\frac{e^2F^2_{1}d^2}{2\pi^4m^2\theta_0^2}\frac{E'^2}{E^2}q_{\min}
    \qquad\qquad\qquad\qquad\qquad\qquad\,\,\nonumber\\
    \cdot\sum^\infty_{n=1}\Theta\!\left(\!n\!-\!\frac{q_{\min}}{|q_-|}\!\right)\!\frac{c^2_n}{n^{4+2\epsilon}}\!\left(\!1\!-\!\frac{2q_{\min}}{n|q_-|}\!+\!\frac{2q^2_{\min}}{n^2|q_-|^2}\!+\!\frac{\omega^2}{2EE'}\!\right)\!,\nonumber\\
\end{eqnarray}
which complies with the coherent bremsstrahlung spectrum in a
straight crystal \cite{coh-bremsstr} (note that the dependence on
$R$ drops out). However, due to the $\theta_0^{-2}$ dependence of
(\ref{rr}), with the increase of $|\theta_0|$ to reach
(\ref{straight-limit}) the radiation intensity attenuates. Besides
that, at large incidence angles the continuous potential
approximation may be invalidated.
\end{description}

\section{Conditions of applicability}\label{sec:conditions}

Our framework in the preceding two sections had been developed by
the principle of maximal theoretical simplicity. In
Sec.~\ref{sec:defl}, in our infinitesimal description of the
particle deflection in the crystal we appealed to the high value of
the particle energy. In Sec.~\ref{sec:rad} we yet adopted the dipole
approximation to radiation emission, which, however, is known
\cite{BKStr} to break down at a sufficiently high energy. Therefore,
we have to investigate whether these two approximations are mutually
consistent under conditions of a real silicon crystal, and if yes,
what is their compatibility domain. Yet, besides the continuous
potential influence on the particle there exists incoherent
scattering on individual nuclei, which affects the particle
deflection as well as radiation. After all, in a case $R\gg |R_c|$
the condition of infinitesimal deflection certainly fails in
vicinity of the volume reflection point, and that may also affect
the radiation spectrum in some frequency domain. The present, last
section comprises estimates of all the mentioned effects.

\subsection{Validity of the infinitesimal deflection approximation}
\begin{figure}
\includegraphics{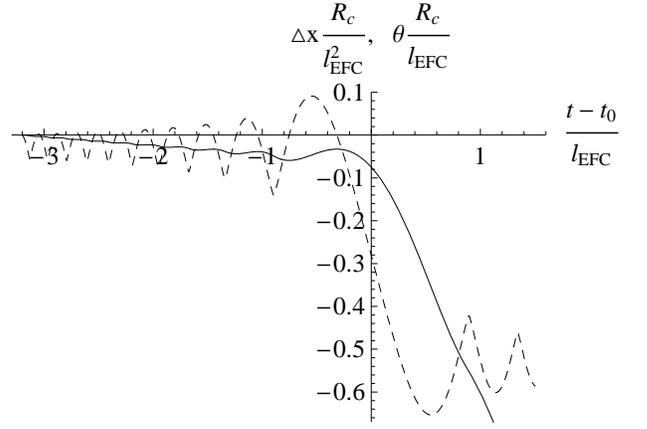}
\caption{\label{fig:deltab} The particle local transverse departure
from the initial straight trajectory $\triangle x(t)$, expressed in
units of $l^2_{\mathrm{EFC}}/R_c$ (solid line), and the local
deflection angle $\theta(t)$, in units of $l_{\mathrm{EFC}}/R_c$
(dashed line), as functions of the distance to the point of tangency
to crystalline bent planes (for some specific impact parameter).
Both curves are built for the room-temperature case $\epsilon=0.4$,
crystal orientation (110) and for $\beta=-0.7$ (see
Eq.~(\ref{beta})).}
\end{figure}

The condition of validity of the straight passage approximation is
the smallness of the  particle transverse displacement relative to
the inter-planar distance. Based on Eq.~(\ref{force-f-L}), let us
evaluate the particle transverse displacement as a function of time:
\begin{eqnarray}\label{double-int}
    \triangle x(t)=\int_{-L/2}^{t}dt'\int_{-L/2}^{t'}dt''\frac{F(t'')}E\qquad\qquad\quad\quad\,\nonumber\\
    =\frac{2}{\pi
    R_c}\sum^\infty_{n=1}\frac{(-1)^nc_n}{n^{1+\epsilon}}\qquad\qquad\qquad\qquad\nonumber\\
    \cdot\int_{-L/2}^{t}dt'\int_{-L/2}^{t'}dt''\sin\left(2\pi
    n\frac{b+\theta_0t''-\xi(t'')}d\right).
\end{eqnarray}
Changing here the order of integrations, and again expanding
$\xi(t'')$ in Taylor series about point $t_0$, one converts the
double integral in (\ref{double-int}) to a single one which is of
Fresnel type:
\begin{eqnarray}\label{deltax}
    \frac{\triangle
    x(t)R_c}{l^2_{\mathrm{EFC}}}=\frac2\pi\sum^\infty_{n=1}\frac{(-1)^nc_n}{n^{1+\epsilon}}\quad\qquad\qquad\qquad\qquad\qquad\nonumber\\
    \cdot\int_{-\frac{L/2+t_0}{l_{\mathrm{EFC}}}}^{\frac{t-t_0}{l_{\mathrm{EFC}}}}d\tau\!\left(\frac{t-t_0}{l_{\mathrm{EFC}}}-\tau\!\right)\!\sin\!\left\{2\pi
    n\!\left(\beta-\tau^2\right)\!\right\}.\nonumber\\
\end{eqnarray}
It behaves as shown in Fig.~\ref{fig:deltab} (by solid line). At
large negative $\frac{t-t_0}{l_{\mathrm{EFC}}}$
\begin{eqnarray}\label{uu}
    \frac{\triangle
    x(t)R_c}{l^2_{\mathrm{EFC}}}\underset{-\frac{t-t_0}{l_{\mathrm{EFC}}}\gg1}\approx\quad\quad\qquad\qquad\qquad\qquad\qquad\qquad\nonumber\\
    \approx\left(1+\frac{t-t_0}{L/2+t_0}\right)\quad\quad\qquad\qquad\qquad\qquad\qquad\qquad\nonumber\\
    \cdot\frac1{2\pi^2}\sum^\infty_{n=1}\frac{(-1)^{n-1}c_n}{n^{2+\epsilon}}\cos\!\left\{\!2\pi
    n\!\left(\beta-\frac{(L/2+t_0)^2}{l^2_{\mathrm{EFC}}}\right)\!\!\right\}\quad\nonumber\\
    +\frac{l^2_{\mathrm{EFC}}}{8\pi^3\!(t\!-\!t_0)^2}\sum_{n=1}^{\infty}\!\frac{(-1)^{n-1}c_n}{n^{3+\epsilon}}\sin\!\left\{\!2\pi
    n\!\left(\!\beta-\frac{(t\!-\!t_0)^2}{l^2_{\mathrm{EFC}}}\right)\!\!\right\}\!.\nonumber\\
\end{eqnarray}
A curious feature here is the weak linear drift at the initial
stage, visualized in Fig.~\ref{fig:deltab} and represented in
Eq.~(\ref{uu}) by the term proportional to
$1+\frac{t-t_0}{L/2+t_0}$. It may be interpreted as a beam
refraction at the entrance to the bent crystal. The refraction angle
sign depends on the impact parameter at the entrance. However, by
the absolute magnitude this effect is small, and for our current
estimates less relevant.

The most relevant is the behavior of function (\ref{deltax}) at
$t>t_0$ where it grows linearly (which corresponds to a motion along
the scattering angle final asymptote), as
\begin{eqnarray}\label{linear}
    \frac{\triangle
    x(t)R_c}{l_{\mathrm{EFC}}}\!\!\underset{\frac{t-t_0}{l_{\mathrm{EFC}}}\gg1}\approx\!\!(t-t_0)\frac{\!\!\sqrt2}\pi\!\sum^\infty_{n=1}\!\frac{(-1)^nc_n}{n^{3/2+\epsilon}}\sin\!\left(\!2\pi n\beta\!-\!\frac\pi4\!\right)\nonumber\\
    +\frac1{2\pi^2}\sum^\infty_{n=1}\frac{(-1)^nc_n}{n^{2+n}}\cos\left\{2\pi n\!\left(\beta-\frac{\left(L/2+t_0\right)^2}{l^2_{\mathrm{EFC}}}\right)\!\!\right\}\qquad\nonumber\\
    +\frac{l^2_{\mathrm{EFC}}}{8\pi^3(t\!-\!t_0)^2}\sum_{n=1}^{\infty}\!\frac{(-1)^{n-1}c_n}{n^{3+\epsilon}}\sin\!\left\{\!2\pi
    n\!\left(\beta-\frac{(t\!-\!t_0)^2}{l^2_{\mathrm{EFC}}}\right)\!\!\right\}\!.\nonumber\\
\end{eqnarray}
In fact, the acting force contribution builds up only before the
particle reaches the final asymptote. For an actual estimate of the
transverse displacement up to that moment one may simply take the
value of (\ref{linear}) at $\frac{t-t_0}{l_{\mathrm{EFC}}}\simeq1$.
For reliability of the straight passage approximation, the
corresponding transverse displacement $\triangle x$ needs to be less
than the inter-planar interval half-width:
\begin{equation}\label{deltax-t0}
    \triangle x(t_0+l_{\mathrm{EFC}})\ll d/2.
\end{equation}
With the use of Eq.~(\ref{linear}), condition (\ref{deltax-t0})
boils down to
\begin{equation}\label{R<<Rc}
    R\ll \frac\pi4 |R_c|.\quad\qquad (\mathrm{for~perturb.~defl.~angle})
\end{equation}
This is physically obvious, since at $R\geq |R_c|$ non-perturbative
effects such as planar channeling, or volume reflection in the bent
crystal already become important.

Yet, for applicability of the stationary phase approximation the
necessary requirement is that the external field coherence length
$l_{\mathrm{EFC}}=\sqrt{2Rd}$ be small compared to the crystal
half-thickness, i. e.,
\begin{equation}\label{RdllL}
    \sqrt{2Rd}\ll L/2.
\end{equation}

In what concerns applicability of the straight passage approximation
to the radiation, the corresponding condition is milder than
(\ref{deltax-t0}, \ref{R<<Rc}). CBBC stems from the oscillatory part
of the particle motion on the initial or the final asymptote,
represented by last lines of Eqs.~(\ref{uu}) and (\ref{linear}).
Denoting that oscillatory part of motion as $\text{\textsf{var}}\,
\triangle x(t)$, the condition for the straight passage
approximation applicability to the description of radiation is
$\text{\textsf{var}}\, \triangle x\big|_{|t-t_0|\sim L/2}\ll d/2$.
Substituting for $\text{\textsf{var}}\, \triangle x(t)$ the last
line of (\ref{uu}) or (\ref{linear}), one arrives at the requirement
\begin{equation}\label{energy-req}
    \frac{L^2}{R^2}\gg\frac{2\sqrt2}{\pi^3}\frac d{|R_c|}.
\end{equation}
That condition is essentially a product of (\ref{R<<Rc}) and
(\ref{RdllL}), so when (\ref{RdllL}) holds very well, inequality
(\ref{R<<Rc}) may be relaxed.

\subsection{Validity of the dipole approximation for radiation}\label{subsec:dip-cond}

Secondly, we had employed dipole approximation for the radiation,
which presumes smallness of the particle deflection angle compared
to the typical radiation angle $\gamma^{-1}$. So, let us evaluate
from Eq.~(\ref{force-f-L}) the \emph{local} angle of deflection from
the straight path. Using again the stationary phase approximation,
one gets
\begin{eqnarray}
  \theta(t) = \int^t_{-L/2}dt'\frac{F(t')}E\quad\qquad\qquad\qquad\qquad\qquad\qquad\quad\nonumber\\
  \approx\frac{l_{\mathrm{EFC}}}{R_c}\frac{2}{\pi}\sum^\infty_{n=1}\frac{(-1)^nc_n}{n^{1+\epsilon}}\int^{\frac{t-t_0}{l_{\mathrm{EFC}}}}_{-\infty}d\tau\sin\left(2\pi
  n\left(\beta-\tau^2\right)\right).\nonumber
\end{eqnarray}
This is an ordinary Fresnel integral; it is observed to converge
within the range $l_{\mathrm{EFC}}$, and its asymptotic forms at
$|t-t_0|\gg l_{\mathrm{EFC}}$ are \footnote{It is evident that
(\ref{theta-away}) is basically a derivative of (\ref{uu}).}
\begin{eqnarray}
  \theta(t)\underset{-\frac{t-t_0}{l_{\mathrm{EFC}}}\gg1}\approx\frac{l^2_{\mathrm{EFC}}}{R_c(t_0-t)}\quad\qquad\qquad\qquad\qquad\qquad\,\nonumber\\
  \cdot\frac1{2\pi^2 }\sum^\infty_{n=1}\!\frac{(-1)^{n-1}c_n}{n^{2+\epsilon}}\cos\!\left(\!2\pi
  n\!\left(\!\beta\!-\!\frac{(t\!-\!t_0)^2}{l^2_{\mathrm{EFC}}}\right)\!\!\right)\!,\,\label{theta-away}
\end{eqnarray}
and
\begin{eqnarray}
  \theta(t)\underset{\frac{t-t_0}{l_{\mathrm{EFC}}}\gg1}\approx\theta_{\mathrm{Born}}\quad\quad\qquad\qquad\qquad\qquad\qquad\qquad\quad\nonumber\\
  +\frac{l^2_{\mathrm{EFC}}}{R_c(t\!-\!t_0)}\frac1{2\pi^2 }\!\sum^\infty_{n=1}\!\!\frac{(-1)^{n-1}c_n\!}{n^{2+\epsilon}}\!\cos\!\left(\!2\pi
  n\!\!\left(\!\beta\!-\!\frac{(t\!-\!t_0)^2}{l^2_{\mathrm{EFC}}}\right)\!\!\right)\!.\nonumber\\
  \label{theta-after}
\end{eqnarray}

Now, for validity in weakly bent crystals of the dipole
approximation to radiation, we need smallness of the oscillatory
part $\text{\textsf{var}}\,\gamma\theta(t)$ at $|t-t_0|\lesssim
L/2$. Substituting in Eq.~(\ref{theta-away}) or the last line of
Eq.~(\ref{theta-after}) $|t-t_0|\to L/2$, and replacing the sum by
its typical value $1/\sqrt2$, we get
\begin{equation}\label{llLR}
    \text{\textsf{var}}\,\gamma\theta\big|_{|t-t_0|\lesssim L/2}\ll1 \quad\Rightarrow \quad \frac
    LR\gg \tilde\theta_V=\frac{2\sqrt2|F_{1}|d}{\pi^2m}.
\end{equation}
Here $\tilde\theta_V$ is a parameter similar to
$\theta_V=\frac{V_0}m$ of \cite{BKStr}. \footnote{For a parabolic
potential, $V_0=|F_{1}|d/4$, which implies
$\tilde\theta_V=\frac{8\sqrt2}{\pi^2}\theta_V\approx1.1\theta_V$.
Thus, the difference between the definitions $\theta_V$ and
$\tilde\theta_V$ is inessential.} With the use of parameters
(\ref{Fd}), numerically one finds
\begin{equation}\label{}
    \tilde\theta_V\approx0.65\cdot10^{-4}
\end{equation}
both for Si (110) and Si (111). It is worth emphasizing that in a
bent crystal the validity of the dipole approximation to radiation
depends on $\tilde\theta_V$ smallness in comparison not with the
Lindhard critical angle
\begin{equation}\label{theta_c}
    \theta_c=\sqrt{\frac d{2|R_c|}}
\end{equation}
(dependent on the particle energy via $R_c$), but with the active
crystallographic plane bending angle $L/R$.

\subsection{Influence of multiple scattering on particle deflection and on radiation}
\subsubsection{Deflection}
The angle of particle deflection in the continuous potential field
also competes with the (rms, plane) angle of multiple scattering.
The latter has a square root dependence on the medium thickness
traversed \cite{Amsler:2008zzb}:
\begin{equation}\label{}
    \theta_{\mathrm{mult}}(\triangle t)=\sqrt{\left\langle\theta^2_{x}\right\rangle_{\mathrm{mult}}}\equiv\sqrt{\frac12\left\langle\theta^2\right\rangle_{\mathrm{mult}}}
    :=\frac1\gamma\sqrt{\frac{\triangle
    t}{l_{\mathrm{mult}}}}\,.
\end{equation}
For \emph{electrons and positrons} in silicon \cite{Amsler:2008zzb}
\begin{equation}\label{}
    l_{\mathrm{mult}}\approx0.13\,\mathrm{mm}.\quad\qquad(e^\pm\,\,\,\mathrm{in\,\,\, Si})
\end{equation}
For multiple scattering not to affect significantly the particle
deflection in the target, $\theta_{\mathrm{mult}}(L)$ must be less
than the angle given by Eq.~(\ref{theta}):
\[
    \sqrt{\frac{L}{l_{\mathrm{mult}}}}\ll\frac{|F_{1}|\sqrt{Rd}}m,
\]
which entails
\begin{equation}
    \frac
    LR\!\ll\!\left(\frac{F_{1}}m\right)^2l_{\mathrm{mult}}d\approx\left\{\begin{array}{c}3.6\\
               2.5 \\
             \end{array}
              \right\}\cdot10^{-2}.\quad \left\{\!\begin{array}{c}
               \mathrm{Si}\,(110) \\
               \mathrm{Si}\,(111) \\
             \end{array}\!
              \right\}\label{yy}
\end{equation}

\subsubsection{Radiation}
Concerning the coherent radiation at typical frequencies, again,
condition (\ref{yy}) is not relevant. Instead, one is to compare
$\theta_{\mathrm{mult}}$ with angle $\text{\textsf{var}}\,\theta(t)$
at $|t-t_0|\sim L/2$. Should we be interested in the radiation
angular distribution, $\theta_{\mathrm{mult}}$ had to be count on
the whole crystal thickness $L$. However, if only the
(angle-integral) radiation spectrum is looked at, for absence of the
multiple scattering influence on it, angle
$\text{\textsf{var}}\,\theta$ at $|t-t_0|\sim L/2$ should be large
compared to the multiple scattering angle only on the length
$l_{\mathrm{EFC}}$:
\begin{equation}\label{}
    \text{\textsf{var}}\,\theta\big|_{|t-t_0|\sim
    L/2}\gg \triangle\theta_{\mathrm{mult}}(l_{\mathrm{EFC}})\, ,
\end{equation}
i. e.,
\[
    \sqrt{\frac{l_{\mathrm{EFC}}}{l_{\mathrm{mult}}}}\ll\tilde\theta_V\frac
    RL.
\]
For the active crystalline plane bending angle this implies
\begin{equation}\label{LRll}
    \frac
    LR\ll\tilde\theta_V\sqrt{\frac{l_{\mathrm{mult}}}{l_{\mathrm{EFC}}}}\approx\left\{\!\begin{array}{c}
               1.7\\
               1.5\\
             \end{array}\!
              \right\}\cdot10^{-4}\!\left(\frac{1\mathrm{m}}R\right)^{1/4}\!\!.\,\, \left\{\!\begin{array}{c}
               \mathrm{Si}\,(110) \\
               \mathrm{Si}\,(111) \\
             \end{array}\!\!
              \right\}
\end{equation}

\subsection{Incoherent bremsstrahlung background}
Still another issue is that the coherent radiation receives a
background from incoherent radiation acts. A standard way to
estimate the incoherent bremsstrahlung intensity in a crystal is to
take the radiation in an amorphous target made of the same material:
\begin{equation}\label{dEBH}
    \frac{d\!E_{\mathrm{BH}}}{d\omega}=\frac{L}{L_0}\left[\frac43\left(1-\frac\omega
    E\right)+\frac{\omega^2}{E^2}\right]\Theta(E-\omega).
\end{equation}
Here $L_0$ is the radiation length, for silicon amounting
\cite{Amsler:2008zzb}
\begin{equation}\label{}
    L_0=9.36~\mathrm{cm}.
\end{equation}
The $\omega$-dependence of (\ref{dEBH}) is mild, and as an estimate
of $dE_{\mathrm{BH}}/d\omega$ one may take its value at
$\omega\simeq0$.

To compare with, the spectral intensity of the CBBC radiation at an
average radiation frequency $\omega\sim\frac{\omega_++\omega_-}4$
(see Figs.~\ref{fig:spectrum}, \ref{fig:spectrum-he}) is about half
of its maximal value (\ref{soft limit}):
\begin{subequations}
\begin{eqnarray}
    \frac{dE_{\mathrm{CBBC}}}{d\omega}\bigg|_{\omega\sim\frac{\omega_++\omega_-}2}\!&\sim&\frac{e^2}{3\pi}\gamma^2\left\langle\theta^2_{\mathrm{Born}}\right\rangle\label{dE0}\\
    &\equiv&\frac{2e^2F^2_{1}Rd}{3\pi^3m^2}\sum^\infty_{n=1}\frac{c_n^2}{n^{3+2\epsilon}}.\qquad\label{dE0a}
\end{eqnarray}
\end{subequations}
\[
    \left(|\theta_0|<L/{2R}\right)
\]
Numerically, Eq.~(\ref{dE0a}) gives
\begin{eqnarray}\label{gamma2theta2}
    \frac{e^2}{3\pi}\gamma^2\left\langle\theta^2_{\mathrm{Born}}\right\rangle&=&\frac{e^2}{3\pi}\left(\frac{F_{1}d}{\pi
    m}\right)^2\frac{2R}d\sum^\infty_{n=1}\frac{c_n^2}{n^{3+2\epsilon}}\nonumber\\
    &\approx&\left\{\!\begin{array}{c}
               4.5\cdot10^{-4} \\
               3\cdot10^{-4} \\
             \end{array}\!
              \right\}\frac R{\mathrm{cm}}.\qquad\left\{\!\begin{array}{c}
               \mathrm{Si}\,(110) \\
               \mathrm{Si}\,(111) \\
             \end{array}\!
              \right\}\qquad
\end{eqnarray}
As had been mentioned at the end of Sec.~\ref{sec:rad}, the coherent
bremsstrahlung spectral intensity is independent of the crystal
thickness $L$.

For CBBC radiation to manifest itself prominently, it must exceed
the incoherent bremsstrahlung contribution:
\begin{equation}\label{dECBBC-larger-dEBH}
    \frac{dE_{\mathrm{CBBC}}}{d\omega}>\frac{dE_{\mathrm{BH}}}{d\omega}.
\end{equation}
With (\ref{gamma2theta2}, \ref{dEBH}), it appears that the ratio
$dE_{\mathrm{CBBC}}/dE_{\mathrm{BH}}$ depends only on the ratio
$L/R$, i. e. on the active plane bending angle, with the
proportionality coefficient
\begin{equation}\label{dEdE}
    \frac{dE_{\mathrm{CBBC}}}{dE_{\mathrm{BH}}}\sim10^{-3}\frac RL.
\end{equation}

\subsection{Radiation at volume reflection (small $\omega$ domain)}
We had mentioned in Sec.~\ref{subsec:dip-cond} that CBBC mechanism
may be responsible for the generation of a large part of the
radiation spectrum even when condition (\ref{R<<Rc}) is violated. In
the latter case, the infinitesimal deflection approximation fails
for evaluation of the particle final deflection angle,
overestimating it, and hence the CBBC formula (\ref{dECBBC-unif})
must overestimate the radiation spectrum at sufficiently small
$\omega$, where it is proportional to the final deflection angle
squared. Let us now estimate the scale of $\omega$ at which
modification of CBBC radiation is needed.

\begin{figure}
\includegraphics[width=85mm]{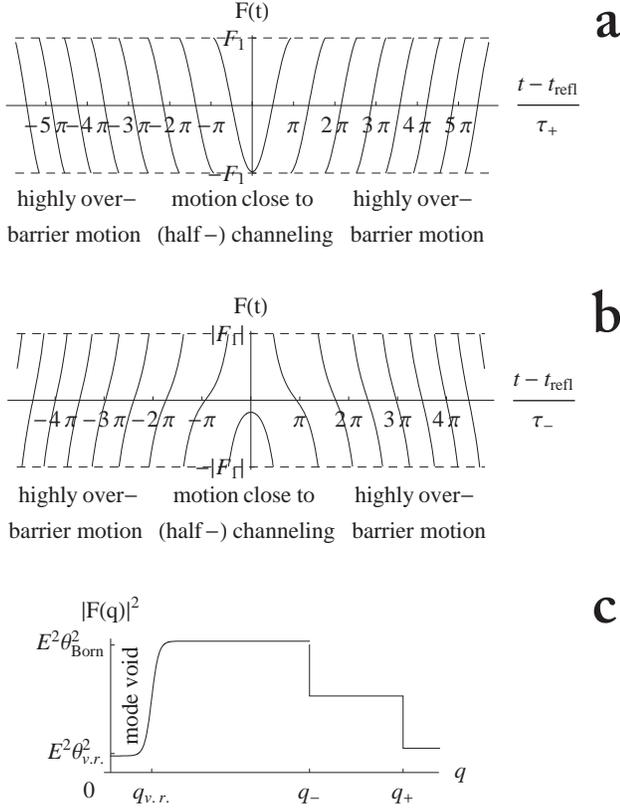}
\caption{\label{fig:Fspectrum} (a) -- exemplary graph of
time-dependence of the force acting in a ((110) oriented,
zero-temperature) bent crystal on a (positively) charged particle
around the volume reflection point, at $R\gg R_c$ (the figure
corresponds to $R=50R_c$). The force discontinuities correspond to
the particle passage through (sharp) potential maxima at the atomic
plane positions. In vicinity of point $t=t_{\mathrm{refl}}$ the
trajectory draws nearly tangential to the maximum potential ridge,
and in that sense is close to (half-) channeling. (b) -- the same
for negatively charged particles. The effective potential maximum is
regular ($F\to0$ on top) and is situated approximately midway the
atomic planes. (c) -- schematic of the force Fourier transform
modulus square. The dominant contribution to $|F(q)|^2$ and
therethrough to $dE_{\mathrm{coh}}/d\omega$ comes from the interval
$q_{\mathrm{v.r.}}\leq q\leq q_+$.}
\end{figure}

\begin{figure}
\includegraphics{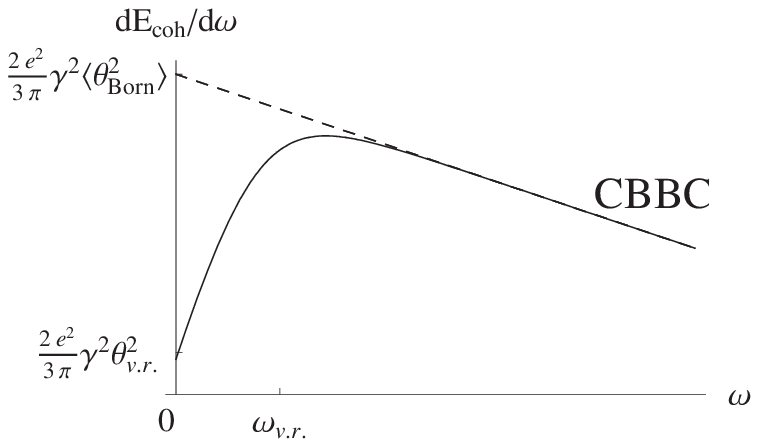}
\caption{\label{fig:v-r} Schematic of a turnover in the coherent
radiation spectrum due to the volume reflection.}
\end{figure}

At $R\gg |R_c|$ the actual mechanism of particle deflection is
volume reflection \cite{Tar-Vor}, whereat the magnitude of the
deflection angle is of the order of Lindhard's critical angle
(\ref{theta_c}). The contributing  $q$-frequencies of particle
oscillation during the volume reflection are effectively
\emph{bounded from below} \footnote{Apart from a small, nearly
constant contribution from the finite total deflection angle.} by
the value equal to \emph{twice} the channeling frequency  $1/\tau$:
\begin{equation}\label{q>}
    q\geq q_{\mathrm{v.r.}}=\frac2\tau ,
\end{equation}
with
\begin{subequations}
\begin{eqnarray}\label{}
    \tau=\tau_+=\sqrt{\frac2{|R_c|d}}\qquad
    (\mathrm{pos.\,char.\,part.})\\
    \tau\simeq\tau_-=\tau_+\frac{\ln\frac{R}{|R_c|}}{2\pi}\qquad
    (\mathrm{neg.\,char.\,part.})
\end{eqnarray}
\end{subequations}
(see Fig.~\ref{fig:Fspectrum}c). The mode void below frequency
(\ref{q>}) arises because in the vicinity of the volume reflection
point $t=t_{\mathrm{refl}}$ (the closest approach to the axis of the
crystal bending) the particle moves in each interval nearly by the
channeling \emph{half} period of the maximal amplitude, crossing the
potential ridges at a nearly grazing angle (see
Figs.~\ref{fig:Fspectrum}a,b) \footnote{The particle also makes one
nearly full undulation period, containing the point
$t=t_{\mathrm{refl}}$, but alone it can not give radiation competing
with coherent radiation from several half-periods. Rather, this
single nearly full period may be regarded as transition from one
semi-channeled motion to another, standing in antiphase, and no
interference of radiation from such antiphased trajectory parts
being possible, despite their identical periods.}. As a consequence,
at radiation frequency
\begin{equation}\label{omega_vr}
    \omega_{\mathrm{v.r.}}=\frac1{\frac1E+\frac1{2\gamma^2q_{\mathrm{v.r.}}}}
\end{equation}
the spectral intensity $dE_{\mathrm{coh}}/d\omega$ related with
$|F(q)|^2$ through Eq.~(\ref{dEdomega-dip}), must have a turnover
(see Fig.~\ref{fig:v-r}), and drop at $\omega\to0$ to
$\frac{2e^2}{3\pi}\gamma^2\theta^2_{\mathrm{v.r.}}$ (with
$\theta^2_{\mathrm{v.r.}}<\left\langle\theta^2_{\mathrm{Born}}\right\rangle$,
see Eq.~(\ref{10}) below) \cite{Landau-Lifshitz}. Hence, at
$\omega\approx\omega_{\mathrm{v.r.}}$ there forms a spectral
maximum, or rather a ``hump" feature, since at
$\omega>\omega_{\mathrm{v.r.}}$ the CBBC spectrum decreases rather
slowly.

A broad maximum similar to the one shown in Fig.~\ref{fig:v-r} had
first been discovered in computer simulations
\cite{Scandale,Chesnokov} of radiation at volume reflection, but its
interpretation was not quite transparent. Now, we may conclude that
as relative to CBBC, the volume reflection effect on radiation is
only of suppressive, not enhancing character. It stems from the
particle inability to sustain in a strong inter-crystalline field a
quasi-periodic motion at too low frequencies -- the over-barrier
particle can not spend in an inter-planar channel a time longer than
the channeling period (actually, half period).

For the perturbative CBBC theory to have a significant applicability
domain, frequency $\omega_{\mathrm{v.r.}}$ must be much lower than
the CBBC spectrum end-point $\omega_+$, which implies
\begin{equation}\label{qvr}
    q_{\mathrm{v.r.}}\ll\frac{\pi L}{Rd},\,\frac m{2\gamma}.
\end{equation}
Condition $q_{\mathrm{v.r.}}\ll\frac m{2\gamma}$ with numerical
values (\ref{Fd}) translates into requirement
\begin{equation}\label{E<10TeV}
    E\ll10\,\mathrm{TeV},
\end{equation}
which is guaranteed by the present accelerator capabilities, whereas
condition $q_{\mathrm{v.r.}}\ll\frac{\pi L}{Rd}$, basically,
coincides with (\ref{energy-req}). The latter may also be regarded
as a condition for the particle energy (see Eq.~(\ref{Egg}) below),
which is, however, not very demanding. So, under the conditions of
dipole radiation volume reflection effects should manifest
themselves in a minor region of the spectrum, indeed.

It is yet relevant to estimate the relative depth of the volume
reflection dip. The actual value of the volume reflection angle (in
crystal with orientation (110)) is
$|\theta_{\mathrm{v.r.}}|\approx\frac\pi2\theta_c$ for positively
charged, and $|\theta_{\mathrm{v.r.}}|\approx\theta_c$ for
negatively charged particles \cite{Bondarenco-v.r}. Therefore,
\begin{eqnarray}\label{10}
    \frac{dE_{\mathrm{v.r.}}(0)}{dE_{\mathrm{CBBC}}(\simeq0)}=\left\{\begin{array}{c}\!\!
                                                                       (\pi/2)^2 \\
                                                                       1
                                                                     \end{array}
    \!\!\right\}\frac{\theta_c^2}{\left\langle\theta^2_{\mathrm{Born}}\right\rangle}\approx\left\{\begin{array}{c}
                                                                       6 \\
                                                                       2.5
                                                                     \end{array}
    \right\}\frac{|R_c|}{R}.\quad\\
    \left\{\begin{array}{c}
                                                                       \mathrm{pos.\, char.\,part.} \\
                                                                       \mathrm{neg.\, char.\,part.}
                                                                     \end{array}
    \right\}\quad\qquad\qquad\nonumber
\end{eqnarray}
Thus, for the dip to be well discernible, one actually needs $R$ to
be at least a few times larger than $R_c$.

\subsection{Crystal and beam optimal
parameters}\label{subsec:optimal}
Let us now assemble conditions (\ref{RdllL}, \ref{energy-req},
\ref{llLR}, \ref{LRll}, \ref{dECBBC-larger-dEBH}) and examine their
mutual compatibility, the variable parameters being $R$ and $L$.
Eqs.~(\ref{llLR}, \ref{LRll}) together read
\begin{equation}\label{}
    \tilde\theta_V\ll\frac LR\ll
    \tilde\theta_V\sqrt{\frac{l_{\mathrm{mult}}}{l_{\mathrm{EFC}}}},
\end{equation}
which imply
\begin{equation}\label{sqrtggg}
    \sqrt{{l_{\mathrm{mult}}}/{l_{\mathrm{EFC}}}}\ggg1.
\end{equation}
This is reminiscent of the Landau-Pomeranchuk-Migdal (LPM) condition
\begin{equation}\label{LPM}
    l_{\mathrm{form}}\ll l_{\mathrm{mult}},
\end{equation}
but at typical radiation frequencies one estimates
$l_{\mathrm{form}}\lesssim2q^{-1}_{\pm}\sim2\frac{Rd}{\pi L}$, and
according to inequality (\ref{RdllL}), it holds that
$l_{\mathrm{form}}\ll l_{\mathrm{EFC}}$. So, the LPM condition
appears to be less crucial than (\ref{sqrtggg}).

For fulfilment of condition (\ref{sqrtggg}), with
$l_{\mathrm{mult}}$ fixed, one needs to have $l_{\mathrm{EFC}}$, i.
e. $R$ and $d$, as low as possible. Note that the value of $d$ is
lower for orientation (110) than for (111), thus orientation (110)
is more beneficial. But as for $R$, at practice it is normally at
least in the range of decimeters, which gives
$\sqrt{{l_{\mathrm{mult}}}/{l_{\mathrm{EFC}}}}\simeq 4$, while at
highest $R\sim10\,\mathrm{m}$ one has
$\sqrt{{l_{\mathrm{mult}}}/{l_{\mathrm{EFC}}}}\simeq1.4$. Thus,
unfortunately, it is impossible to demand inequality (\ref{sqrtggg})
as really strong. Anyway, the optimal value for the active
crystallographic plane bending angle is about
\begin{equation}\label{optimal}
\frac
LR\sim\tilde\theta_V\left(\frac{l_{\mathrm{mult}}}{l_{\mathrm{EFC}}}\right)^{1/4}\sim
1.3\cdot10^{-4}. \,\qquad (\mathrm{optimal})\quad
\end{equation}
Then, the parameter of radiation non-dipoleness (also known as
$\rho$-parameter \cite{BKStr}) is
\begin{equation}\label{gammatheta05}
    \gamma\theta \big|_{|t-t_0|\sim L/2}\sim\tilde\theta_V\frac RL\simeq0.5,
\end{equation}
whereas the parameter of radiation decoherence due to multiple
scattering is about the same:
\begin{equation}\label{}
    \frac{\triangle\theta_{\mathrm{mult}}(l_{\mathrm{EFC}})}{\theta|_{|t-t_0|\sim
    L/2}}\sim\frac
    R{\tilde\theta_VL}\sqrt{\frac{l_{\mathrm{EFC}}}{l_{\mathrm{mult}}}}\approx0.5.
\end{equation}

In view of the narrowness of condition (\ref{sqrtggg}), it seems
reasonable to suggest that since the size of the coherence length
remains the same for all locations within a uniformly bent crystal,
and the length $l_{\mathrm{mult}}$ is constant as well, then even if
condition (\ref{LRll}) fails (multiple scattering effects are
substantial), the spectrum \emph{shape} may still be roughly
described by the present theory, only the intensity being suppressed
by a factor depending on the ratio
$l_{\mathrm{EFC}}/l_{\mathrm{mult}}$. However, evaluation of such a
factor is beyond the scope of the present article.

Within our framework, presuming all the abovementioned conditions to
be fulfilled, let us check the last crucial condition
(\ref{dECBBC-larger-dEBH}). With (\ref{optimal}), ratio (\ref{dEdE})
will amount $dE_{\mathrm{CBBC}}/dE_{\mathrm{BH}}\sim7$, which is
satisfactorily high.

Other relevant conditions (\ref{RdllL}, \ref{energy-req}) are rather
easy to fulfil. At bending angle (\ref{optimal}) Eq.~(\ref{RdllL})
demands for the crystal thickness
\begin{equation}\label{Lgg}
    \frac Ld\gg \frac{8R}{L}\approx 6\cdot10^4\quad\Rightarrow\quad L\gg \left\{\begin{array}{c}
                                                                       12 \\
                                                                       20
                                                                     \end{array}
    \right\}\mu\mathrm{m}.\quad \left\{\begin{array}{c}
                                                                       \mathrm{Si}\,(110) \\
                                                                       \mathrm{Si}\,(111)
                                                                     \end{array}
    \right\}
\end{equation}
Eq.~(\ref{energy-req}) together with (\ref{optimal}) sets the lower
bound for the electron energy:
\begin{equation}\label{Egg}
    E\gg\frac{2\sqrt2}{\pi^3}\frac {R^2}{L^2}|F_{1}|d\sim1\,\mathrm{GeV}.
\end{equation}

Towards experimental investigation of CBBC itself, let us yet
determine the full set of parameters at which CBBC effects are least
deteriorated and the sharpest. As we had mentioned, lower $R$ are
favored for cleaner CBBC, but $R$ is tightly interrelated with $L$,
which must satisfy inequality (\ref{Lgg}). Taking, marginally
\begin{equation}\label{Lstar}
    L_\star\approx\left\{\begin{array}{c}
                                                                       50 \\
                                                                       80
                                                                     \end{array}
    \right\}\,\mu\mathrm{m} \qquad \left\{\begin{array}{c}
                                                                       \mathrm{Si}\,(110) \\
                                                                       \mathrm{Si}\,(111)
                                                                     \end{array}
    \right\}
\end{equation}
(such thin samples are available -- cf. Ref.~1 of \cite{Guidi}),
entails through Eq.~(\ref{optimal})
\begin{equation}\label{Rstar}
    R_\star\approx\left\{\begin{array}{c}
                                                                       0.4 \\
                                                                       0.6
                                                                     \end{array}
    \right\}\,\mathrm{m}.\qquad \left\{\begin{array}{c}
                                                                       \mathrm{Si}\,(110) \\
                                                                       \mathrm{Si}\,(111)
                                                                     \end{array}
    \right\}
\end{equation}
For what concerns the electron energy, to avoid a pronounced volume
reflection turnover one needs, according to (\ref{10}), smaller
ratio $R/|R_c|$, i. e. higher $E$. Besides that, if one wants the
``ankle" features in the radiation spectrum not to be smeared out by
the photon recoil effects (cf. Figs.~\ref{fig:spectrum} and
\ref{fig:spectrum-he}), one should arrange the condition $\frac{2\pi
L\gamma^2}{Rd}\ll E$, which under (\ref{optimal}) implies
\begin{equation}\label{Ell}
    E\ll\frac{m^2d}{2\pi}\frac RL\approx\left\{\begin{array}{c}
                                                 300 \\
                                                 500
                                               \end{array}
    \right\}\mathrm{GeV}.\quad\left\{\begin{array}{c}
                                                 \mathrm{Si}\,(110) \\
                                                 \mathrm{Si}\,(111)
                                               \end{array}
    \right\}
\end{equation}
If in marginal fulfilment of (\ref{Ell}) one takes
\begin{equation}\label{Estar}
    E_\star\approx150\,\mathrm{GeV},
\end{equation}
we derive
\begin{equation}\label{}
    |R_c|_\star\sim \left\{\begin{array}{c}
                                                 25 \\
                                                 37
                                               \end{array}
    \right\}\,\mathrm{cm},\quad\left\{\begin{array}{c}
                                                 \mathrm{Si}\,(110) \\
                                                 \mathrm{Si}\,(111)
                                               \end{array}
    \right\}
\end{equation}
whereby $R_\star/|R_c|_\star\approx1.6$, and according to
Eq.~(\ref{10}), the dip should not develop. Parameters (\ref{Lstar},
\ref{Rstar}, \ref{Estar}) are ``ideal" for checking the calculations
in the present paper; the spectrum thereat must look like that in
Fig.~(\ref{fig:spectrum}). The $\omega$ range is up to
$\frac1{\frac1E+\frac{Rd}{2\pi\gamma^2L}}\sim35\div50\,\mathrm{GeV}$
and the expected spectral intensity maximum is
$\frac{dE_{\mathrm{CBBC}}(0)}{d\omega}\approx 0.04$. The most
demanding condition seems to be the initial electron beam
collimation degree $\delta\theta_0<\frac
L{2R}\approx70\,\mu\mathrm{rad}$. If not achieved, an averaging of
the radiation spectrum over the electron beam incidence angles must
be performed.

\section{Summary and discussion}\label{sec:summary}
The present study substantiates the notion that spectral intensity
of radiation from ultra-high-energy electrons and positrons in a
bent crystal
is a sum of contributions from particle motion intervals on which
the local angle between the particle velocity and the bent
crystalline planes is definite, and so each such contribution is
similar to coherent bremsstrahlung in a straight crystal. The length
of an elementary coherence interval is $\simeq
l_{\mathrm{EFC}}=\sqrt{2Rd}$, implying that (i) the crystal must be
much thicker than the mentioned value (see Eq.~(\ref{RdllL})); (ii)
the radiation spectral intensity
(Eqs.~(\ref{dECBBC-gen}-\ref{dECBBC-unif})) is proportional to
$l^2_{\mathrm{EFC}}\propto Rd$, and does not depend on the crystal
thickness. Still, under the ``moderately high energy" condition
(\ref{soft-rad-cond}) the crystal thickness determines the spectrum
extent, and therethrough, the total energy emitted.

The characteristic feature of CBBC from a single electron is the
well-defined end of the radiation spectrum, whose position depends
on the active crystallographic plane bending angle $L/R$. At an
angle of electron incidence on the crystal comparable to $\frac
L{2R}$ this end of spectrum splits into a pair of breaks (see
Fig.~\ref{fig:spectrum}). That feature must in principle be
experimentally verifiable with a sufficiently well collimated
initial beam ($\delta\theta_0<\frac L{2R}$). The best experimental
conditions for investigating CBBC were described in
Sec.~\ref{subsec:optimal}.

We have also qualitatively discussed the modification of the
coherent radiation spectrum in the domain of small $\omega$ owing to
the onset of the volume reflection phenomenon possible when $R\gg
|R_c|$. That modification is of purely suppressive character and
manifests itself as a dip at the beginning of the spectrum. Next to
the dip, around frequency (\ref{omega_vr}) there appears to be a
maximum in the spectrum, but it is not to be interpreted as a
resonance.


The theoretical description adopted in this article had resorted to
many simplifications -- it did not properly incorporate the
temperature dependence of the potential, neglected multiple
scattering, and relied on an infinitesimal approximation to the
particle deflection (in the bulk of the medium) as well as on dipole
description of the radiation. Conditions (\ref{RdllL},
\ref{optimal}-\ref{Egg}) under which those approximations hold,
altogether appear to be restrictive for the crystal bending angle
(see Eq.~(\ref{optimal})), so generalization to a non-dipole
treatment, and an account of the multiple scattering would be highly
desirable. Nonetheless, let us mention that the dipole CBBC
conditions are quite nicely met, e. g., in recent experiment
\cite{Scandale}. Comparison of the CBBC theory with the available
experimental data is intended elsewhere.

In conclusion, let us remark that although our paper presumed
dependence of the crystal deformation only on one, longitudinal,
coordinate, in principle higher-dimensional deformation cases are
conceivable, emerging under application of torsion, or owing to
intrinsic crystal mosaicity. In those cases the stationary phase
approximation must still be applicable, but the description should
inevitably become more sophisticated.

\textbf{Acknowledgement.} The author wishes to thank A.~V.~Shchagin
for fruitful discussions.


\begin{thebibliography}{00}

\bibitem{Afonin}
A.~G.~Afonin \emph{et al.}, JETP Lett. \textbf{88} (2008) 414.
\bibitem{Scandale}
W.~Scandale \emph{et al.}, Phys. Rev. A \textbf{79} (2009) 012903.
\bibitem{Tar-Vor}
A.~M. Taratin and S.~A.~Vorobiev, NIMB \textbf{26} (1987) 512.
\bibitem{Chesnokov}
Yu.~A.~Chesnokov, V.~I.~Kotov, V.~A.~Maisheev, and I.~A.~Yazynin,
JINST \textbf{3} (2008) P02005.
\bibitem{Arutyunov}
V.~A.~Arutyunov, N.~A.~Kudryashov, V.~M.~Samconov, and
M.~N.~Strikhanov, Nucl. Phys. B \textbf{363} (1991) 283.
\bibitem{coh-bremsstr}
G.~Diambrini Palazzi, Rev. Mod. Phys. \textbf{40} (1968) 611;
M.~L.~Ter-Mikayelyan. High Energy Electromagnetic Processes in
Condensed Media, Wiley, New York, 1972.
\bibitem{Solovyov}
V.~G.~Baryshevsky, I.~Ya.~Dubovskaya, and A.~O.~Grubich, Phys. Lett.
A \textbf{77} (1980) 61; V.~V.~Kaplin, S.~V.~Plotnikov, and
S.~A.~Vorob'ev, Zh. Tekh. Fiz. \textbf{50} (1980) 1079; S.~Belucci
\emph{et al.}, Phys. Rev. ST \textbf{7} (2004) 023501; A.~V.~Korol,
A.~V.~Solov'yov, and W.~Greiner, Int. J. Mod. Phys. E \textbf{13}
(2004) 867; N.~F.~Shul'ga, V.~V.~Boyko, and A.~S.~Esaulov, Phys.
Lett. A \textbf{372} (2008) 2065.
\bibitem{cryst-und}
S.~Bellucci \emph{et al.}, Phys. Rev. Lett. \textbf{90} (2003)
034801.
\bibitem{Ivanov}
Yu.~M. Ivanov \emph{et al.}, JETP Lett. \textbf{81} (2005) 99.
\bibitem{Guidi}
V.~Guidi, A.~Mazzolari, D.~De Salvador, and A.~Carnera, J.~Phys. D
\textbf{42} (2009) 182005; S.~G.~Lekhnitskii. Theory of Elasticity
of an Anisotropic Body. Mir, Paris, 1981.
\bibitem{Lindhard}
J.~Lindhard, Mat. fys. medd. Kgl. Danske vid. Selskab. \textbf{34}
(1965) 14.
\bibitem{Biryukov-UFN}
V.~M.~Biryukov, Yu.~A.~Chesnokov, and V.~I.~Kotov, Sov. Phys. Usp.
\textbf{37} (1994) 937.
\bibitem{Tsyganov}
E.~N.~Tsyganov, Fermilab Report No. TM-682, 1976 (unpublished);
Fermilab Report No. TM-684, 1976 (unpublished).
\bibitem{Olver}
F.~W.~J.~Olver. Asymptotics and Special Functions. Academic Press,
New York, 1974.
\bibitem{Apostol}
T.~M.~Apostol. Introduction to Analytic Number Theory. Springer, New
York, 1976.
\bibitem{Landau-Lifshitz}
L.~D.~Landau and E.~M.~Lifshitz. The Classical Theory of Fields.
Pergamon, London, 1962.
\bibitem{BKStr}
V.~N.~Baier, V.~M.~Katkov, and V.~M.~Strakhovenko. Electromagnetic
processes at high energies in oriented single crystals. World
Scientific, Singapore, 1998.
\bibitem{Amsler:2008zzb}
B. Rossi. High Energy Particles. Prentice-Hall, New York, 1952;
  C.~Amsler {\it et al.}  [Particle Data Group],
  Phys. Lett.  B \textbf{667} (2008) 1.
\bibitem{Bondarenco-v.r}
M.~V.~Bondarenco, arXiv:0911.0107.
\end{thebibliography}
\end{document}